# Domain Transfer in Latent Space (DTLS) Wins on Image Super-Resolution – a Non-Denoising Model

Chun-Chuen Hui[1,2]    Wan-Chi Siu[1,2]    Ngai-Fong Law[2]
[1]Caritas Institute of Higher Education    [2]The Hong Kong Polytechnic University
cchui@cihe.edu.hk  enwcsiu@polyu.edu.hk  ngai.fong.law@polyu.edu.hk

## Abstract

*Large scale image super-resolution is a challenging computer vision task, since vast information is missing in a highly degraded image, say for example for scale ×16 super-resolution. Diffusion models are used successfully in recent years in extreme super-resolution applications, in which Gaussian noise is used as a means to form a latent photo-realistic space, and acts as a link between the space of latent vectors and the latent photo-realistic space. There are quite a few sophisticated mathematical derivations on mapping the statistics of Gaussian noises making Diffusion Models successful. In this paper we propose a simple approach which gets away from using Gaussian noise but adopts some basic structures of diffusion models for efficient image super-resolution. Essentially, we propose a DNN to perform domain transfer between neighbor domains, which can learn the differences in statistical properties to facilitate gradual interpolation with results of reasonable quality. Further quality improvement is achieved by conditioning the domain transfer with reference to the input LR image. Experimental results show that our method outperforms not only state-of-the-art large scale super resolution models, but also the current diffusion models for image super-resolution. The approach can readily be extended to other image-to-image tasks, such as image enlightening, inpainting, denoising, etc.*

## 1. Introduction

Diffusion models have rapidly advanced image synthesis in recent years, achieving outstanding performance [4, 5, 11, 12, 14, 24–26, 28, 29, 32, 35, 38]. Denoising Diffusion Probabilistic Model (DDPM) [9] was firstly proposed to generate images by reversing a Markovian process. This is to remove Gaussian noises gradually from a latent pattern via a learned noise prediction network, typically U-Net [11, 26] or U-ViT adopted in [2].

Figure 1a shows the training procedure of the DDPM. To train the U-Net in DDPM, an input image $x_0$ (where 0 indicates $t$=0, in the noiseless domain) is added with noise $N$ according to timestep $t$, which is randomly drawn from 1 to

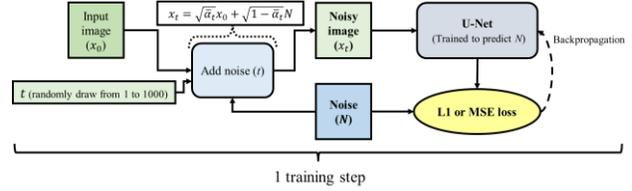

(a) Training step of Diffusion model (DDPM)

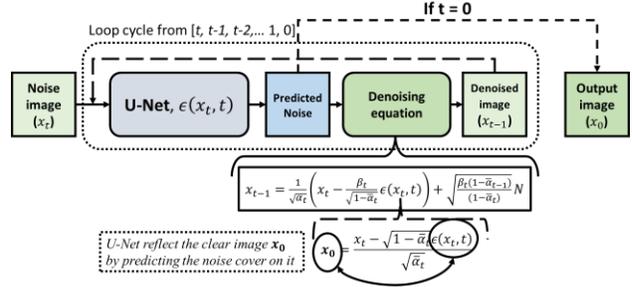

(b) Sampling steps of Diffusion model (DDPM)

Figure 1. Training and image generation flow of DDPM

1000. The noise adding process follows eqn.1 with scaling $\sqrt{\bar{\alpha}_t}$ as shown below,

$$x_t = \sqrt{\bar{\alpha}_t} x_0 + \sqrt{1 - \bar{\alpha}_t} N \qquad (1)$$

The noisy image $x_t$ is inputted into the U-Net model to predict the noise $N$ added to $x_0$, where the prediction is used to calculate the loss for image generation. Mean absolute error (L1 loss) or mean square error (MSE loss) can be used as a loss function in the training procedure. By this training structure, the U-Net can learn the statistical properties of the image dataset, $\{x_0\}$, in training. Note that the training and generation procedures can identify image distributions from different timestep $t$, since the training is generalized by using random timesteps.

For the generation process as shown in Fig. 1b, a noisy image $x_t$ is inputted into a loop cycle. In the loop, the trained U-Net firstly produces a noise pattern for $x_t$. The predicted noise pattern is then processed with a denoising equation and the denoised image $x_{t-1}$ is then outputted. The denoised image $x_{t-1}$ goes back to the beginning of the loop and to make denoising again with the U-Net again until *t*=0. At *t*=0, the prediction of the U-Net is supposed noiseless, thus, the denoising equation is no longer employed. This last prediction result becomes the output i.e., image $x_0$.

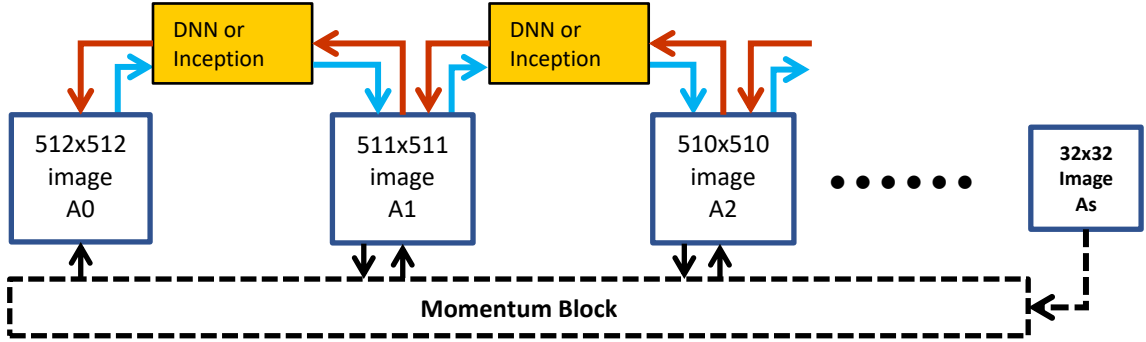

Figure 2. Conceptual idea of our model

Note that in DDPM, the denoise equation can usually expressed as,

$$x_{t-1} = \frac{1}{\sqrt{\alpha_t}}\left(x_t - \frac{\beta_t}{\sqrt{1-\bar{\alpha}_t}}\epsilon(x_t, t)\right) + \sqrt{\frac{\beta_t(1-\bar{\alpha}_{t-1})}{(1-\bar{\alpha}_t)}}N(0,1)$$

(2)

where $\epsilon(x_t, t)$ represents the output from learnt U-Net, and is used to predict noise. We want to clarify that although the U-Net is predicting the noise pattern, it is statistically linked to the possible noiseless image $x_0$. By reverting equation 1, we can obtain,

$$x_0 = (x_t - \sqrt{1-\bar{\alpha}_t}N)/\sqrt{\bar{\alpha}_t} \quad (3)$$

This means that the predicted noise image always refers to the latent real image representation of $x_0$. The diffusion model makes use of the trained U-Net to predict hidden clean image $x_0$ and makes use of the denoising equation to guide the transformation process from latent space to photo-realistic space. We can consider that noise is just a medium throughout the generation process.

SR3 [30] proposed a diffusion super-resolution model which has similar evaluation and training procedures with DDPM. The U-Net in SR3 was redesigned to use a low-resolution (LR) image as guidance to the noise prediction. SR3 concatenates the LR image with Gaussian noise at first step and uses intermediate denoising result in every denoising step, where 2000 steps in total is default in SR3 or IDM [7] models. This conditioning system confirms the pattern predicted by the U-Net which can match the representation of the LR image. Hence, SR3 can generate a super-resolution (SR) image consistent with the LR image. These image transformation diffusion models still bear much the idea of image generation rather than transforming a small image to a high-resolution image [6, 7, 15, 18, 19, 30, 31, 33, 36, 37].

We consider that adding Gaussian noises is just a means

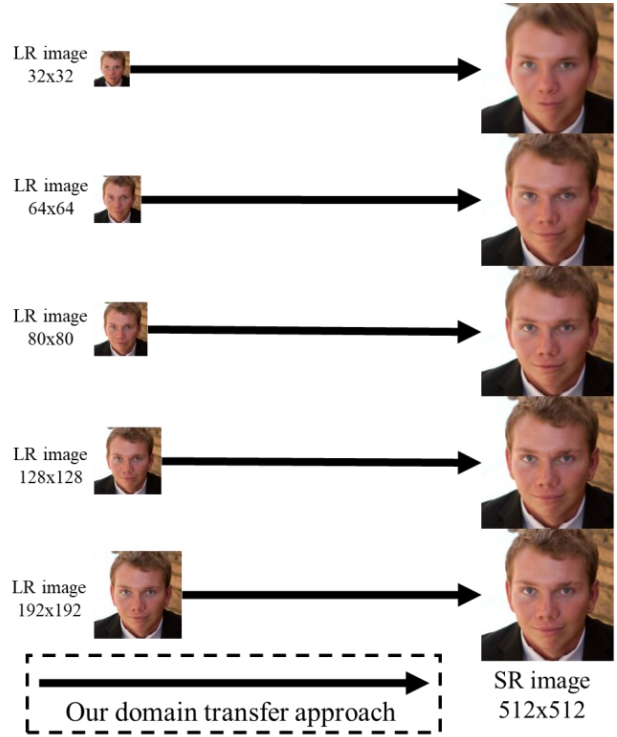

Figure 3. Image super-resolution with our domain transfer method. It can process various input image sizes with only one trained model.

to build a photo-realistic high-resolution image space with reference to an initial noise pattern, and the training procedure is to let the U-net learn the statistical properties of the training dataset and the statistical differences between time t-stages. While the denoising procedure of diffusion models is to make use of the predicted statistical properties obtained from the trained U-Net (and the derived formulation) to move toward the constructed photo-realistic high-resolution image space. Therefore, we propose our super-resolution approach which shares some concepts of diffusion models but without the use of Gaussian noise at

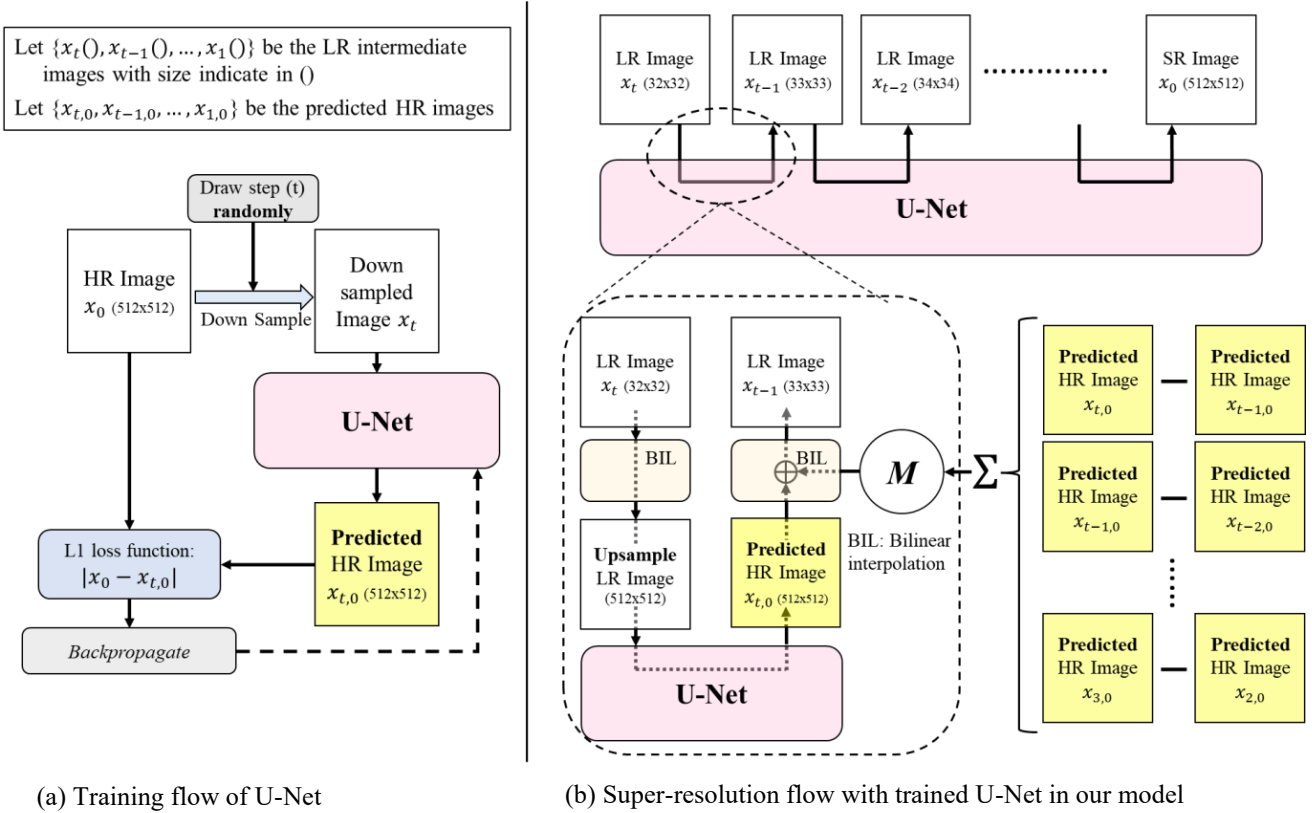

(a) Training flow of U-Net  (b) Super-resolution flow with trained U-Net in our model

Figure 4. Training and image super-resolution flow of our model

all. We make use of the statistical properties predicted from the trained DNN (for realization just used a trained U-Net) to gradually transfer a LR image to a SR image as shown in Fig.2. Image generation or image super-resolution is done gradually via each scale step as shown in the diagram (indicated by the red arrows).

As verified by our experimental work, just making use of the consecutive reverse process is able to produce super-resoluted images with reasonable quality, but the quality is inferior to those produced by Diffusion Models [7, 30]. It is because a Markov chain reverse process may sideway the path which leads to the loss of fidelity. Hence, we relate each step with the LR image in question and introduce an element $M$, where $M$ acts like an opposite momentum to errors. The super-resolution results are substantially better than the state-of-the art approaches, including Diffusion Models. Details are found below and in our ablation studies.

## 2. Methodology

In this work, we introduce our novel Domain Transfer in Latent Space (DTLS) framework for image super-resolution based on the concept of iterative domain transfer as shown in Fig. 2. One key advantage of our iterative approach is the flexibility it provides in both input and output resolutions.

As shown in Fig. 3, DTLS can accommodate varying input resolutions, including uncommon sizes like 80×80 or 192×192 pixels. A second advantage of DTLS is its ability to perform extremely large-scale transformations, as it will be demonstrated in our experimental results. In this paper, we utilize face super-resolution as an example application to showcase the effectiveness of our proposed DTLS method.

### 2.1. Theory

One similar concept between diffusion model and our design is that, both models learn sets of image space from degraded domain to clear domain. The difference is, the degrade domain of diffusion model is achieved by Gaussian noise and our design aims to handle noise (the difference) between various levels of natural image degradation. Diffusion models are carefully designed systems that break down noise adding and denoising into a number of steps, for which they are well-defined with the probabilistic models.

The U-Net actually occupies a big part in the diffusion model, for example, a better design of the U-Net can enhance the performance of image synthesis by the diffusion model [5]. As the U-Net is trained to build a structure for producing the latent image space at various stages via Gaussian noise, it eventually can produce high-resolution image in the final high-resolution latent domain (making use of the trained parameters). We assume that the U-Net can also be trained and be used to produce intermediate domain results and eventually lead to the final result, with a high quality super-resoluted image without involving the mechanism of noise operations. We refer it to as a domain transfer, which transfers a 249x249 image to a 248x248 image say for example, and its reverse for the reverse processing. The transfer has to be small and gradual such that the U-Net is able to accommodate. The fact making this possible is that, as shown in Fig. 2, we have the HR(512x512) and LR(32x32) image pairs available, and in the training, we force (with well-designed loss-functions) to form the weights in the U-net to fulfill our objective to produce high quality HR image in the reverse process.

## 2.2. Overall Structure

---
**Super-resolution flow**

   **Input:** $x_{t=T}$
   $M = 0$
   **for** $t = T, T - 1, …, 1$ **do**
      $x_{t,0} \leftarrow U(x_t, t)$
      **if** $t \neq T$ **then**
         $M = M + x_{t,0} - x_{t-1,0}$
      $x_{t-1} \leftarrow D(x_{t-1,0}, t-1) + D(M, t)$
   **return** $x_0$

---
**Training procedure**

   **repeat:**
      $t \sim$ Randomly sample from $\{1…T\}$
      $x_0 \sim$ Randomly sampling from **Dataset** $\{x_0\}$
      $x_t \leftarrow D(x_0, t)$
      $x_{t,0} \leftarrow U(x_t, t)$
      $Loss \leftarrow |x_{t,0} - x_0|$
      **Backpropagate**
   **Until break**

---

Figure 4a shows the training procedure of our model. Similar to diffusion models, the training image $x_0$ is randomly down sampled to $x_t$. The U-Net is used to predict a HR image $x_{t,0}$ with any down sampled $x_t$. A L1 loss function takes $x_0$ and $x_{t,0}$ to calculate the error to perform backpropagation. By randomly down sample HR images, our U-Net learns the statistical properties of all images $\{x_0\}$ in all degraded image space, from the smallest (ex. 32×32) to the largest (ex. 512×512).

For the super-resolution part as shown as Fig. 4b, the LR image is gradually enlarged with the concept of domain transfer. To achieve domain transfer, the model requires a clear image as guidance, which is the predicted HR image $x_{t,0}$.

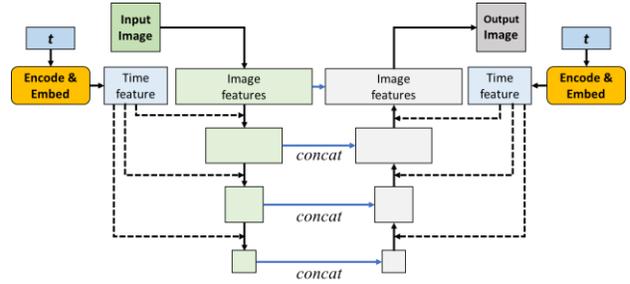

Figure 5. Overall structure of U-Net used in our DTLS, note that *t* was inputted for guiding the U-Net to produce correct restoration result.

To make use of the input image of smaller size for the U-Net, we firstly apply bilinear interpolation to up sample the LR image to the size of HR image. For instance, as shown in Fig. 3b, the LR image (32×32) is firstly up sampled to 512×512 before inputting to the U-Net. After the prediction, a down sampling with bilinear interpolation will transfer the large image (sixe:512×512) to $x_{t-1}$ for which the size is 33×33.

Only using this structure, the approach does not work too well. Hence in between the upsampling action and after making the prediction from the U-Net, an element *M* which is the sum of all previous errors is added (see Section 3.2). This is similar to back projection design [9, 20, 21, 34]. The summation of these inversion errors can be expressed as $\sum_{s=T}^{t}(x_{s,0} - x_{s-1,0})$, where *T* is the total transformation steps and *t* is the current step. Note that at first and second steps, since $x_{t,0}$ and $x_{t-1,0}$ still have not predicted, thus *M* is equal to zero at these first two steps.

We also denote the training procedure and super-resolution flow in pseudo codes as shown at the beginning of this section, where *D()* is the down sample operator which includes a down sample and a up sample bilinear interpolation. The purpose of up sample is to make all input images the same size for U-Net operations. The U-Net is represented by *U()*, which aims to predict the HR image $x_{t,0}$. The overall structure of the U-Net is shown as in Fig. 5. Similar to the one in the diffusion model, the U-Net uses timestep *t* to condition the down sampling and upsampling blocks.

## 2.3. Error and Momentum

The above procedure is a Markov chain process, meaning that the trnsfer is between two neighbouring domains, but not other domains. However, the error will accumulate beteen pairs of domains in the transfer process, which utimately leads to big error at the final stage and to the final resultant image. Fig.6 shows that, for the first iteration, the U-Net generates a predicted HR image that could slightly deviate from the expected value. When we move further towards the ground truth image $x_0$, bigger errors are formed due the accumulation. Note that the CNN model was not trained for distorted inputs. It will produce even bigger errors with distorted inputs. This initializes a vicious cycle where the error $E_t$ accumulates all coming iterations.

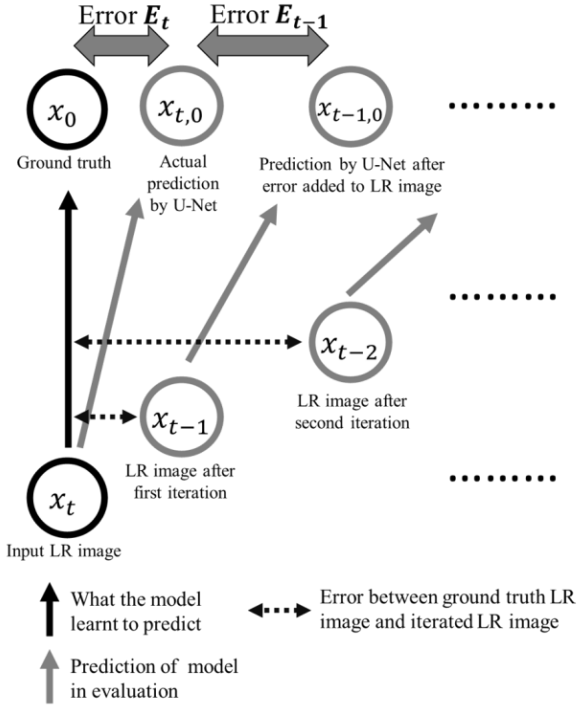

Figure 6. Flow diagram to explain how error accumulate during domain transfer.

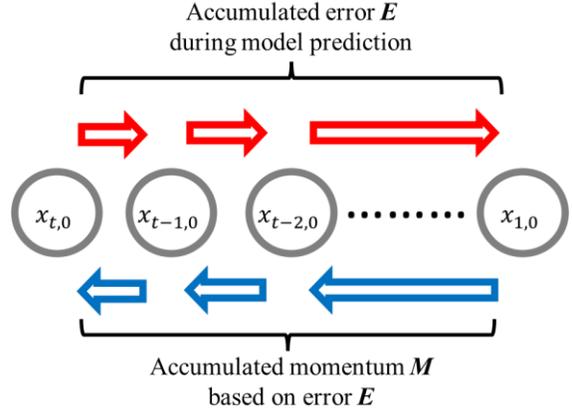

Figure 7. While error $E$ appear between from $t$ to $t$-1, our momentum $M$ can counter in opposite direction.

To address the issue of error accumulation in iterative models, we propose a momentum term $M$. As the prediction error $E$ accumulates over iterations, we simply add momentum $M$ in the opposite direction. As shown in Fig. 4 and Fig. 7, $M$ is calculated as the sum of previous errors. The change from $x_{t,0}$ to $x_{t-1,0}$ can be obtained by $(x_{t-1,0} - x_{t,0})$ where the change here is considered as error. Therefore, to acquire contrary direction of error, a swap action is taken as $(x_{t,0} - x_{t-1,0})$ and thus $M$ is summation as $\sum_{s=T}^{t}(x_{s,0} - x_{s-1,0})$.

The physical meaning of momentum is further explained in Fig. 7, as the prediction errors tend to propagate to the model's outputs in the same direction over time (the red arrow moving right), momentum works to cancel out these errors by building up in the opposite direction (the blue arrow moving left). Through this cancellation effect at each iteration, momentum helps control the growth of total error. Additional ablation studies on the impact of momentum are presented in section 3.3. Incorporating this momentum term enables our model to iteratively transform images while mitigating accumulated error artifacts.

## 3. Experimental Results

**Dataset** To test our DTLS system in face super-resolution, we used FFHQ 1024x1024 dataset which contains 70,000 images [17]. The training image is randomly cropped to the targeted size. Horizontal flip is also used as image augmentation for robustness training.

**Implementation details** We implemented our design in Py-Torch 1.10.2 with the training on a single NVIDIA RTX3090 (24G VRAM). We chose two common large super-resolution scales for the experiment. The first one is from 16×16 to 128×128, where we applied the U-Net with channel widths of {64, 128, 256, 1024}. The other super-resolution set is 32×32 to 512×512, and the channel widths are {32, 64, 128, 128, 256, 256, 512}. Optimizer AdamW with betas {0.9, 0.999} and eps 1e$^{-8}$ were used, and the learning rate was set to 2e$^{-5}$. 50K iterations with batch size of 32 and 374K iterations with batch size of 4 were applied on 16 to 128 and 32 to 512 respectively.

### 3.1. Comparisons with Non-Diffusion Models

While DTLS has the advantage of handling variable input and output resolutions, most state-of-the-art face super-resolution models were designed for single-scale transformations. Therefore, to conduct comparison, we tried the evaluation on three scales: 32×32 to 512×512, 64×64 to 512×512 and 128×128 to 512×512. Two image quality assessment metrics SSIM and PSNR are used for quantitative comparisons. Tab. 1 shows the comparison of SSIM and PSNR of our approach with two state-of-the-art methods, Face-SPARNet [3] and GCFSR [10]. Face-SPARNet makes use of spatial attention network to perform face super-resolution. GCFSR designs a scale-controllable encoder-generator structure to perform face super-resolution in any enlargement scale. To ensure a fair comparison, we downloaded and used their pre-trained network provided on their project websites with dataset of FFHQ. Visual examples for comparison are also shown in Fig. 8. Our model outperforms both Face-SPARNet and GCFSRs in all super-resolution scales on SSIM and PSNR. This is especially true in extreme scales like 32 to 512 (×16). I.e., when performing large-scale super-resolution, our DTLS is able to stably and smoothly upsample faces while minimizing artifacts. In contrast, Face-SPARNet and GCFSR tended to either over-sharpen details or produced under-resolved outputs in these extreme transformation settings. At smaller enlargement scales, with faces closer to ground truth resolution, DTLS still consistently generated images with clear facial details. Overall, DTLS has demonstrated robust and stable performance compared to existing models, especially for the challenging task of high-magnitude super-resolution where originally the accumulation of artifacts poses a significant problem for iterative approaches.

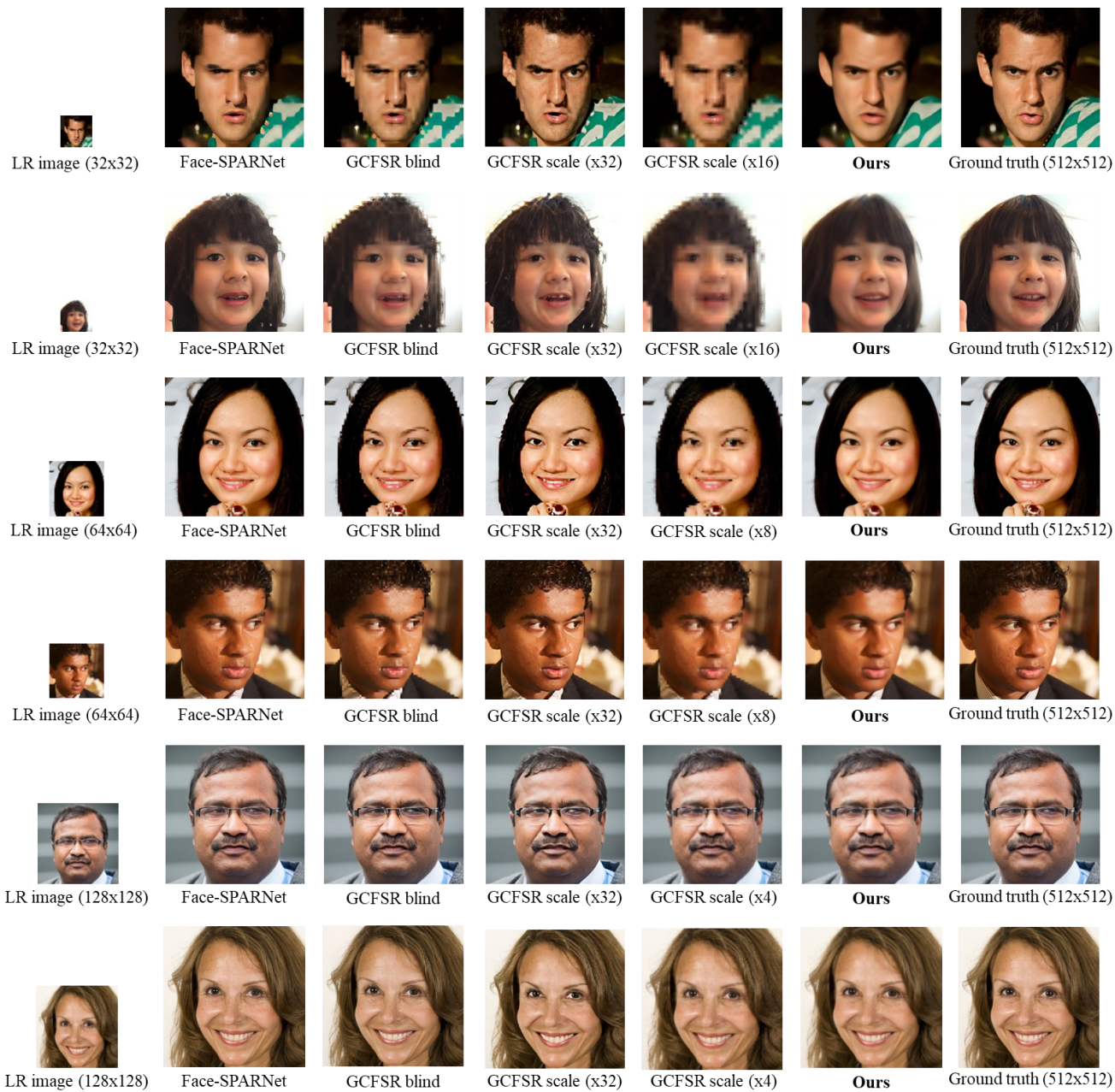

Figure 8. Visual comparison with other SOTA non-diffusion models.

| Method (SSIM↑ / PSNR↑) | 32→512 | 64→512 | 128→512 |
|---|---|---|---|
| Face-SPARNet | 0.6044 / 21.9249 | 0.7288 / 26.1935 | 0.8179 / 29.2282 |
| GCFSR Blind Face Restoration | 0.6071 / 21.6363 | 0.7270 / 25.7160 | 0.8455 / 29.9659 |
| GCFSR (×32) | 0.5818 / 21.2648 | 0.6581 / 24.2048 | 0.7282 / 26.5061 |
| GCFSR (×16, ×8, ×4) | 0.6431 / 22.8709 | 0.7528 / 26.8418 | 0.8747 / 31.3928 |
| **Ours** | **0.7252 / 24.7641** | **0.8141 / 28.6682** | **0.8921 / 32.4579** |

Table 1. Metrics comparison with other SOTA non-diffusion models. Our method performs better on all scales.

## 3.2. Comparisons with Diffusion Models

To enable comparison between DTLS and diffusion models which perform super-resolution by conditioning the denoising process on Gaussian noise, we focused our evaluation on the two scales where pre-trained diffusion model checkpoints were readily available:

1) 16×16 to 128×128 and 2) 32×32 to 128×128. Since the original publications and model websites for SR3 and IDM only provided pre-trained weights applicable to these resolution ranges, we confined our analysis to evaluating and visualizing results within the same input-output domains for an unbiased comparison. This has limited our tests to relatively small magnification factors, but it still allows us to have a good comparison of our DTLS system with domain transfer approach to the noise-conditioned generation methods employed by diffusion models for face super-resolution.

As shown in Fig. 9, for up sampling from 16×16 to 128×128 (an enlargement factor of ×8), the DTLS continues to stably generate super-resolved results that closely match the structures in the ground truth images. While SR3 and IDM produce high-fidelity images, their outputs do not align as well with the ground truth HR targets compared to DTLS, as evidenced by low SSIM and PSNR scores in Tab. 2. For the 32×32 to 128×128 (×4) case, SR3 is not able to sufficiently leverage the LR image as condition for the guidance and often produced distorted facial features. DTLS more accurately transfers between domains by incrementally shifting through intermediate resolutions, rather than relying on a single noise-conditioned generation step. This allows it to better preserve structural fidelity to the target domain when performing large-scale face super-resolution. While SR3 produces high-quality images in the 16 → 128 case, it is important to note that diffusion-based models like SR3 generates a new image guided by noise rather than explicitly enlarging a degraded input. As discussed previously, this means that the SR outputs from diffusion models may not perfectly produce the target HR images, even they result in photo realistic images. This is evident in Fig. 9, where SR3's outputs have high fidelity but do not precisely align with the ground truths. On quantitative consistency as shown in Tab. 3, DTLS has obtained images with better consistency and been able to produce SR images that correspond to the LR inputs.

| Method (SSIM↑ / PSNR↑) | 16→128 | 32→128 |
|---|---|---|
| SR3 | 0.6922 / 23.5806 | 0.4890 / 18.0262 |
| IDM | 0.6426 / 23.0934 | 0.7465 / 24.0079 |
| **Ours** | **0.7367 / 24.0065** | **0.8636 / 27.6423** |

Table 2. Metrics comparison (SSIM/PSNR) with diffusion super-resolution models. Our method performs better on all scales.

| Method (Consistency ↓) | 16→128 | 32→128 |
|---|---|---|
| SR3 | 0.394 | 2.439 |
| IDM | 0.442 | 1.597 |
| **Ours** | **0.125** | **0.648** |

Table 3. Comparison of consistency, consistency measures MSE (×10$^{-5}$) between the LR inputs and the down-sampled SR outputs.

## 3.1. Ablation Studies

In this section, we focus more on the novelty, momentum $M$. We have explained the concept of $M$ in Fig. 6, a further exploration is conducted in the following section.

When we firstly applied our iterative image enlargement design, the quality of SR results was not too good, and there were many parts with over-smoothing. We found that the quality of predicted HR image $x_{t,0}$ from U-Net drops with more iterations. Various evidence from our experimental works have shown us the effect in Fig. 6, where the error of prediction accumulated throughout the domain transformation process. This is the reason why we have introduced $M$, as a stabilizer.

To study the momentum $M$, we have compared various usages of $M$, say on 32 → 512 face super-resolution. Note that momentum $M$ does not obtain by training the U-Net. Table 4 shows the results of checking the usefulness of M, where all results were obtained from one pre-trained U-Net. Metrics were measured in SSIM and PSNR. Let us discuss cases with labels (i) to (vi) as shown in Table 4. (i) Without the use of momentum $M$ to reduce restoration errors, it performs the worst among all other situations that make use of M properly. (ii) If we used $M$ for two consecutive levels only, which means $M$ no longer accumulates long term previous errors, it only gains 0.003 and 0.067 in SSIM and PSNR and it is still better than those without the use of momentum, but now with almost neglectable improvement. (iii) Adding $M$ in the predicted HR domain image $x_{t,0}$ instead of $x_t$, (in the LR domain) which indicates that $M$ is no longer needed to be bilinear degraded before (refer to Fig. 4b, $M$ is passing through the bilinear interpolation block): that is instead, $M$ can be added to $x_{t,0}$ since errors are calculated in HR domain ($x_{t,0} - x_{t-1,0}$). However, our study shows that this method has a decrease in quality (-0.0029 SSIM) and (-0.0133 PSNR) compared with our current strategy on the use of momentum. This is because the use of the $M$ after the U-Net prediction cannot let the U-Net to predict a better $x_{t,0}$ with ($x_t + M$). (iv) This is our current strategy. Our current tactic allows the U-Net to predict the HR image with modulated LR image from the use of $M$. (v) Lastly, we may degrade $M$ before putting $M$ in the LR image $x_t$. As mentioned in (iii), previous prediction errors are calculated in HR images, a HR image cannot be added into a LR image without down sampling, otherwise it will distort the domain of the LR image and make DTLS have difficult to make domain transfer properly. This can be seen that both SSIM and PSNR of (v) have decreased massively. Meanwhile, the visual comparison of (iv) and (v) is shown in Fig. 9. The result of adopting $M$ to our DTLS without down sampling gives huge distortion because the domain transferring process is interrupted.

| Method | SSIM | PSNR |
|---|---|---|
| (i) w/o $M$ | 0.7110 | 24.5093 |
| (ii) $M$ w/o accumulation | 0.7141 | 24.5760 |
| (iii) $M$ adding in HR domain | 0.7233 | 24.7508 |
| **(iv) Our current design** | **0.7252** | **24.7641** |
| (v) w/o degrading $M$ | 0.0832 | 15.9781 |

Table 4. Metrics comparison of our DTLS in different usages of momentum $M$.

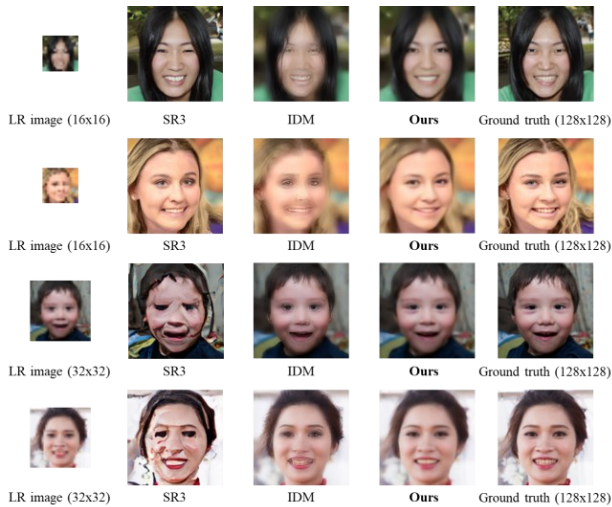

Figure 9. Visual comparison with diffusion faces super-resolution models.

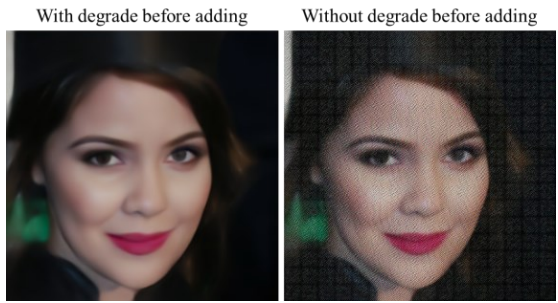

Figure 10. Without degradation of momentum $M$, the SR result distorted massively (right) compared to proper usage of M (left).

## 4. Conclusion

In this work, we introduced our model on Domain Transfer in Latent space (DTLS) for super-resolution imaging, which demonstrates the potential for noiseless iterative design for image-to-image translation tasks. Leveraging the concept of domain shifting across intermediate representations and guided by an error-correction momentum term $M$, DTLS successfully overcomes issues of error accumulation. In our experiments on face super-resolution, DTLS outperformed state-of-the-art methods across all scales of magnification in super-resolution imaging, validating its ability to handle highly flexible, multi-scale transformations.

Super-resolution is suggested in this paper as it is a commonly ill-posed problem in image-to-image transformation. The exploration of our model should not end here. Our proposed novel approach could also be extended to other image transformation tasks. We recommend tasks such as low-light image enhancement, deblurring, denoising, inpainting, etc. The goal is for DTLS to serve as a versatile image transformation architecture applicable to a wide variety of restoration and enhancement problems through further research.

More broadly, our framework opens new opportunities beyond existing approaches by enabling general any-input to any-output flexibility. For instance, it may allow freely adjusting image properties like brightness over a continuous range for low-light enhancement. We hope our work inspires further research on domain transfer paradigms. With continued development, iterative domain transfer models may become a universally applicable approach for image transformation tasks. Our model presents just a starting point, and more advances are still needed to unlock the full promise of this new direction.

ACKNOWLEDGMENT

This work is partly supported by the Caritas Institute of Higher Education (1SG200206) and UGC Grant (UGC/FDS11/E05/22) of the Hong Kong Special Administrative Region.

# Appendix

This appendix includes further analysis of our Domain Transfer in Latent Space (DTLS) model with more visual examples of our experimental results. You can find the source code of DTLS in https://github.com/GreyCC/DTLS.

**A.** Model design

**A1.) Backbone model** We have used U-Net as our backbone model in transferring image from one domain to another domain, which is similar to diffusion models [5-7, 11, 12, 15, 19, 26, 28-

32, 36, 37] and non-denoising diffusion models [1, 27]. U-Net design can achieve image super- resolution but not the optimized choice over some state-of-the-art network designs, such as [13, 20, 22, 23]. The reason for continuing to apply U-Net is because the current design of U-Net can handle various of timestep better. Let us refer to fig. 5. The encoded and embedded time feature *t* is utilized in both image encoding and decoding processes. While in each level of the encode and decode processes, the U-Net concatenates image features from encoding side to decoding side. This design allows *t* to guide the restoration from extracted features to the output image with both time feature and the concatenation of preliminary image features which are also infected by time feature, resulting an optimal use of the time feature. The capacity to identify time step is especially important in a multiple-steps model, or otherwise the recall process will be inconsistent. Therefore, we have found that the U-shape design is very suitable for both diffusion and non-denoising iterative models.

**A2.) Error accumulation** We have mentioned that errors from prediction of the U-Net accumulate throughout the domain transfer process in section 2.3 on our model, which is different from that of the denoising diffusion models. As for denoising diffusion models, the image for processing is completely noisy at the beginning. The U-Net of these models were trained and aims at producing the right amount of noise. This is achieved by making use of the noise predicted from the U-Net and the denoising equation. This is the reason that the denoising diffusion models can obtain less error in the beginning of influencing, and errors are not easy to accumulate to affect the final result. In our approach, for a quick realization, we may use a step of domain transfer of $2^n$ where n is an integer, say from 64 to 128 ($2^6$ to $2^7$) or even 16 to 128 ($2^4$ to $2^7$). In the latter case, the error accumulation is even larger. Hence, we have to use a Momentum block which acts as a stabilizer to obtain results of really good quality.

**A3.) Momentum destinations** Momentum is designed to cancel errors generated from the predicting process of the U-Net. Note that the difficulty of prediction is less when we come closer to the final result. To enhance the performance of domain transfer, momentum is used to ensure the U-Net can go into the right path of the prediction. Without using momentum, our experimental results show a distance away from the ground truth image. Overall, the use of momentum can stabilize the domain transfer process. A visual comparison of using or not using the momentum block, *M,* is shown as in Fig. A1.

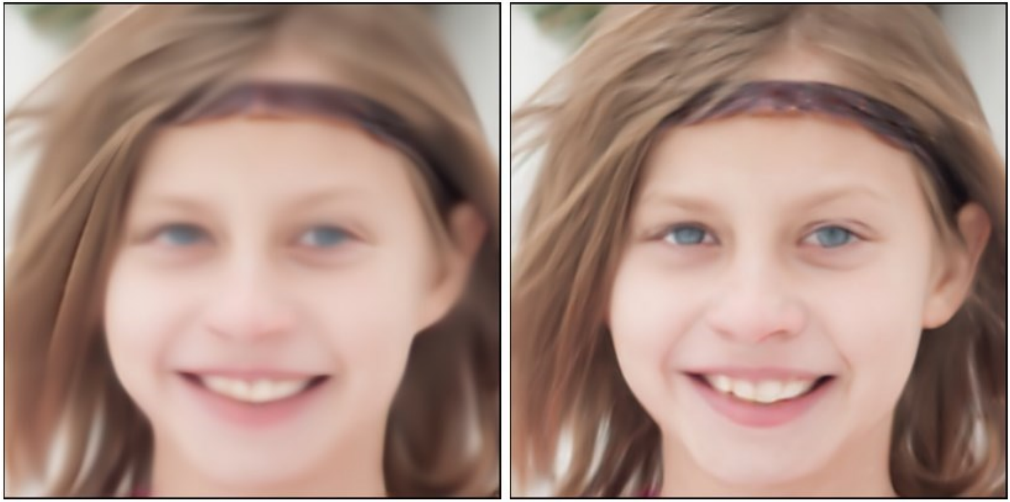

Figure A1. Without momentum (left) and after using of momentum (right)

**B.** Training analysis

B1.) L1 or MSE loss? In denoising diffusion models, most authors suggested using MSE loss as the loss function for the training procedure. However, we suggest to use L1 loss as our loss function in our approach and there are two major reasons. First, MSE loss is not accurate enough for domain transfer. As introduced by DDPM [11], the U-Net in denoising diffusion models aims to predict noise where the noise pattern is more obvious in latent representations. This makes the training of the U-Net for DDPM easier than our DTLS. While our model requires the U-Net to predict across domain layers, which is more challenging and thus a loss function which can obtain better accuracy is necessary. Second, as we are testing on image super-resolution, although much information is lost due to the size of the image in question is small, the LR image still shares the same structure with the HR image. While in denoising diffusion model, the closer to the photo-realistic domain, the less noisy the intermediate image is. However, there is a large difference between the predicted and ground truth images and the MSE loss can help comprehend such issues. However, DTLS does not require MSE loss to handle situations like those in denoising diffusion models, and L1 loss is the most suitable loss function for training the U-Net. Fig. B1 shows the properties of L1 and MSE losses, where x-axis represents the error between prediction and ground truth while y-axis represents the calculated loss after the loss function. The transition points are at -1 to 1, while within [-1,1] the MSE loss trends to give less amount of loss which will reduce the learning power of the model. While if the error is out of the range of [-1,1], the MSE loss gives way larger loss than the L1 loss, this fits the second point mentioned above that DDPM needs MSE loss to train the U-Net to predict noise at timestep, t, is small. Table B1 shows the quantitative performance between the use of L1 loss and the use of MSE loss in the training stage. It is obvious that L1 loss outperforms MSE loss in PSNR and SSIM consistently.

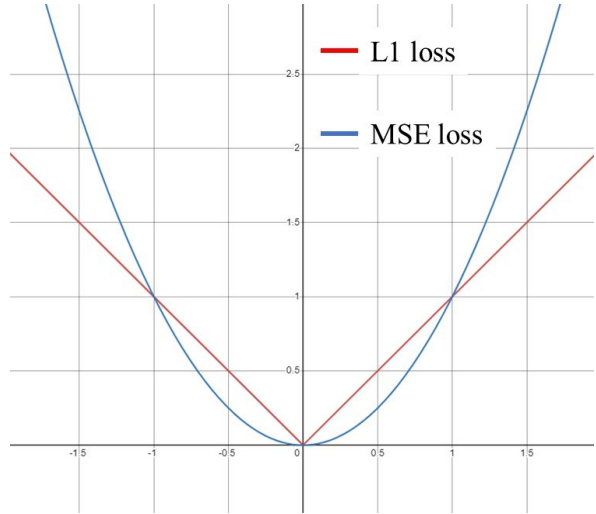

Figure B1. Differences between L1 loss and MSE loss

| 16   128 | PSNR↑ | SSIM↑ | Consistency↓ |
|---|---|---|---|
| **L1 Loss** | **24.0065** | **0.7367** | **0.12523** |
| MSE Loss | 23.9653 | 0.7274 | 0.14142 |

Table B1. Comparison of using L1 loss and MSE loss to train our model. Experiment done on 16 to 128 with the stride of 4.

**B2.) Predict $x_0$ or $x_{t-1}$?** As discussed in A2, the prediction error is larger with larger scale difference between two domains. If we follow this concept, we should design the model and train the U-Net to predict $x_{t-1}$ from $x_t$ gradually, instead of referencing to $x_0$. However, in our inferencing approach, we suggest referencing back to $x_0$. We have several reasons to do so. First, we have to apply our newly designed momentum as a tool to let our process to go directly into the space of the final domain, which is at t=0. To predict $x_{t-1}$ from $x_t$, we cannot simply use the trained U-net, but we have to use it with respect to the space of the final destination, i.e. $x_0$. This is the most crucial reason that we are not just using the U-net for domain transform, $x_t \rightarrow$ $x_{t-1}$. Second, it is due to training difficulty of predicting $x_{t-1}$ in the required steps. In our training, we have to train for allowing 30 intermediate domains at the same time. In this case, as we said, we have to take reference to $x_0$ domain as well. While in the case of predicting $x_{t-1}$ with $x_t$ alone, it is required to make domain transfer for over 30 pairs of domains, and the model have to accommodate this complex situation. This hardens the process of training. Third, although the disparity of $x_0$ and $x_t$ is larger than $x_{t-1}$ and $x_t$, this in fact helps the training of the U-Net. This is due to the loss and backpropagation. With small loss in the training, the network trends not to learn effectively. Hence, the design of moving towards to $x_0$ helps the accuracy of predicting in domain transfer processes. Figure B2 shows the comparison of designing the model in predicting $x_{t-1}$ with our current DTLS design. It shows that domain transfer with only the U-Net for predicting $x_{t-1}$ cannot obtain results as good as those by using an additional momentum block.

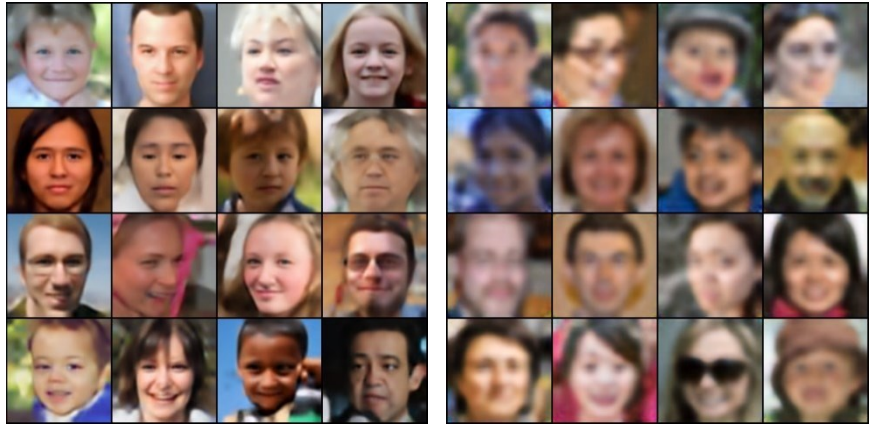

Figure B2. Intermediate training result at 20K iterations (16 to 128). Left is our current DTLS design and right is trained to predict $x_{t-1}$

**B3.) Total steps of domain transfer?** In our experimental part, we have stated that the setting of strides for 32 to 512 and 16 to 128 are respectively 16 and 4. Thus, the total steps of domain transfer for these two settings are 30 and 28 respectively. The reason for setting stride of such is because we found that the total steps of the DTLS model affect the final quality of super-resoluted results. The reason for this phenomenon resonates with our discussion in the last paragraph. That is the total pairs of images the U-Net needs to learn. With smaller strides, more steps are needed and thus more pairs of image representation are memorized by the U-Net during the training stage. However, more pairs mean harder to train for producing a well-tuned U-Net. The visual comparison of stride of 16 in (32 to 512) SR and stride of 1 in (32 to 512) SR is shown in fig. B3. Note that the quality of super-resoluted images with stride 1 is much worse than our current setting: with stride 16 and a total of 30 steps. Nonetheless, we have found that we cannot use larger stride for our model as this will lose the size flexibility of our super- resolution model.

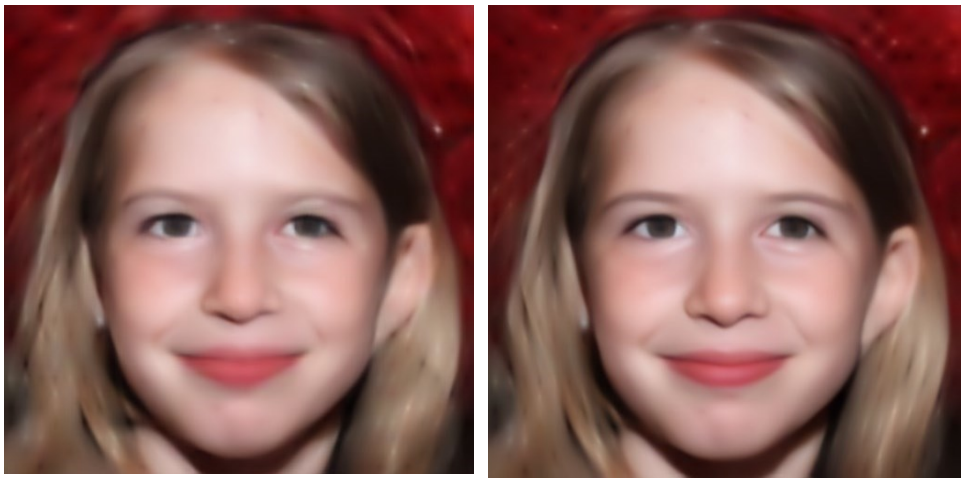

Figure B3. Result comparison of stride of 1 (left) and stride of 16 (right) of 32 to 512 super-resolution scale

## C. Related works

Since it is lack of enough space in the main body of the paper, let us add back the reviews of some of the related works in literature.

**C1.) Diffusion Super-Resolution Models** Except for SR3 [30] and IDM [7], there are more diffusion super-resolution models that have been published [18, 19, 31, 22, 37]. We did not compare these papers in the main session of our paper as they have similar concept with SR3 and IDM.

**ResShift** [37]: Unlike SR3 and IDM that concatenate on LR image to guide the diffusion to generate a fine detail image, ResShift proposed a denoising process to gradually fill in the residual between HR and LR images also with concatenation of LR image for the U-Net. The forward process of ResShift has been redesigned to contain residual $e_0$, where $e_0 = y_0 - x_0$ and $x_0$ is ground truth HR image and $y_0$ is the down sampled $x_0$. The formulation of forward process of ResShift is as follows:

$$q(x_t|x_{t-1}, y_0) = \mathcal{N}(x_t; x_{t-1} + \alpha_t e_0, \kappa^2 \alpha_t I), t = 1, 2, \ldots, T$$

where $\kappa$ is the hyper-parameter for tuning the noise variance. The U-Net of ResShift can be denoted as $f_\theta(x_t, y_0, t)$ and it is trained to predict $x_0$ instead of noise pattern. Since the forward process included part of the residual between $x_0$ and $y_0$, the prediction of the U-Net assumed that the latent space of the Gaussian noise contains the residual as well and thus ResShift can transit $y_0$ to $x_0$ gradually with shifting the residual $e_0$ iteratively.

**ResDiff** [31]: This is a two stages method which applies both CNN network and diffusion model to generate SR images. ResDiff first applies a pre-trained CNN model to enlarge LR image to form an intermediate SR image which is lack of high-frequency texture. The enlarged image is then proceeded for the diffusion model to predict high-frequency content of the image. This is said that SR3, IDM and ResShift make use of interpolated LR image to guide the prediction of the U-Net, but ResDiff uses CNN to enlarge the LR image to guide the denoising iterations. The U-Net and the denoising process are still similar to DDPM where it is designed to generate realistic-likely image. However, ResDiff added a frequency-domain splitter to separate intermediate result to focus on high-frequency domain. Thus, the result after denoising is high-frequency only SR image which will be combined with the intermediate SR image predicted from the pre-trained CNN structure and form the final SR result.

**SRDiff** [19]: Similar to ResShift, SRDiff defines a transition from LR to HR gradually in the denoising process. The diffusion model of SRDiff obtains residual between LR and HR image during the reversing steps with the use of the U-Net as the noise predicting unit. The major difference between SRDiff and ResShift is that SRDiff make use an image encoder to encode LR image to guide the U-Net for noise prediction instead of concatenating to the Gaussian noise.

**StableSR** [33]: StableSR implements latent diffusion model (LDM) [28] to generate SR image. Since LDM encodes the noisy image with a pre-trained encoder before the noise prediction with the use of U-Net, they cannot concatenate LR image directly with the Gaussian noise unless a new training of the encoder is taken as action. In fact, StableSR obtained another image encoder to generate layers of image features to guide in the U-Net's layers. This is similar to guide timestep $t$ in the U-Net.

**DWTrans** [18]: Pretrained LDM was also applied as the backbone structure of DWTrans, where they proposed the use of an extra window adjustable transformer (WAT) after the prediction of the LDM. The prediction of the LDM and the LR image is encoded again for the WAT to fuse LR image into LDM prediction. The WAT is designed to acquire extra global feature from the LR image unlike non-adjustable transformer which can only extract local features. After the fusion of the WAT, the fused image feature is decoded to form the final SR result.

**C2.) Non-denoising diffusion models** Except for our proposal, the concept of not using noise in diffusion models is also available and there are a few published works on non-denoising models.

**Heat Dissipation Model (HDM)** [27]: HDM designed a non-noise iterative system by using heat dissipation as forward process with trained U-Net as reversing model. To perform heat dissipation, HDM fades out high frequency spectrum on Discrete Cosine Transformed image, where the effect of the fading is like blurring the image. To reverse from highly blurred image to photo-realistic image, HDM applies the U-Net model that is trained to predict the previous step $x_{t-1}$ to gradually reverse and generate high frequency information. However, as we have also discussed in section **B2**, predicting $x_{t-1}$ does not give enough loss to train the model at backpropagation. Thus, HDM still proposed to add a little amount of noise in training and for evaluation process.

**Cold Diffusion Model** (CDM) [1]: CDM designed a generative model which can use arbitrary format of degradation do and can perform reversion, for applications like: denoising, inpainting, super-resolution, deblurring, etc. CDM also applies U-Net as their backbone model to achieve prediction. The U-Net is trained to predict $x_0$ and the reverse process of CDM makes use of residual between predicted $x_0$ and down-sampled $x_0$, $x'_{t-1}$. Formulation of the reverse step of CDM can be denoted as, $x_{t-1} = x_t + x_0 - x'_{t-1}$. However, similar to HDM, CDM is required to add a little amount of Gaussian noise to enhance the performance of generated images.

**D.** Additional visual results from our approach (DTLS)

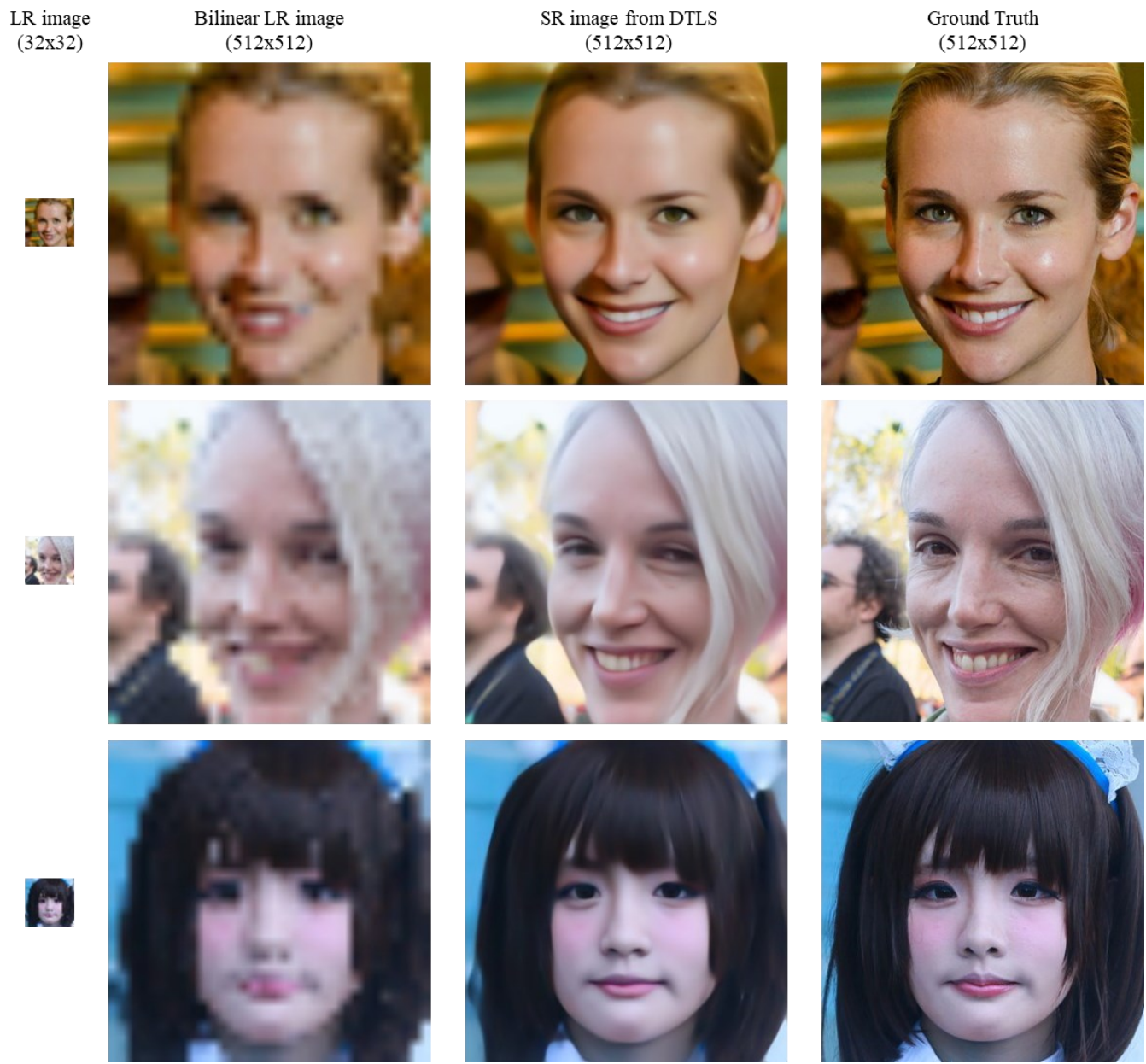

Figure D1. Visual examples for large SR: 32 to 512

| LR image (32x32) | Bilinear LR image (512x512) | SR image from DTLS (512x512) | Ground Truth (512x512) |
|---|---|---|---|
| 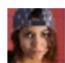 | 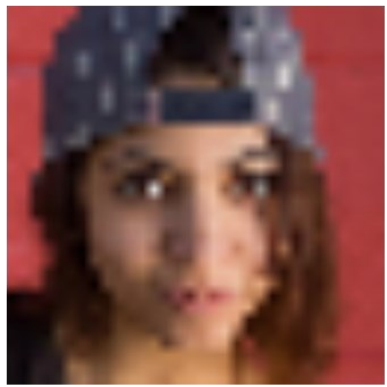 | 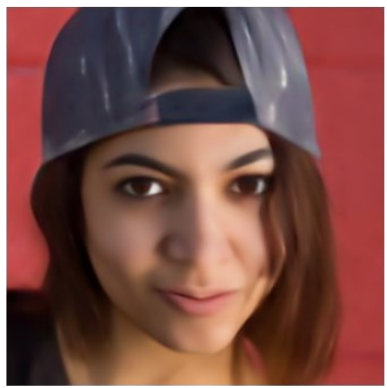 | 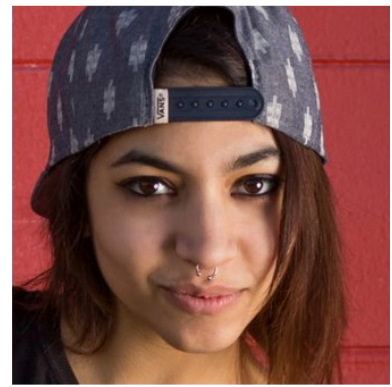 |
| 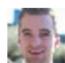 | 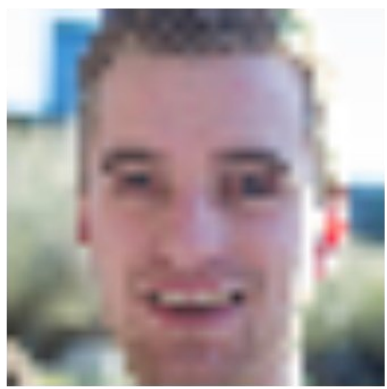 | 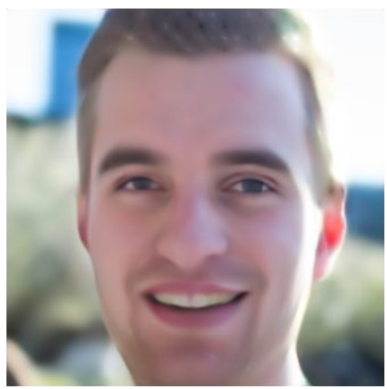 | 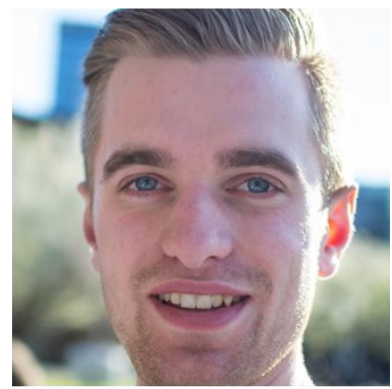 |
| 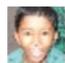 | 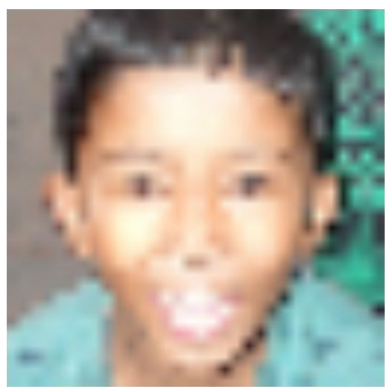 | 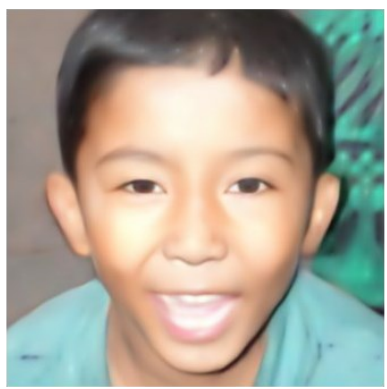 | 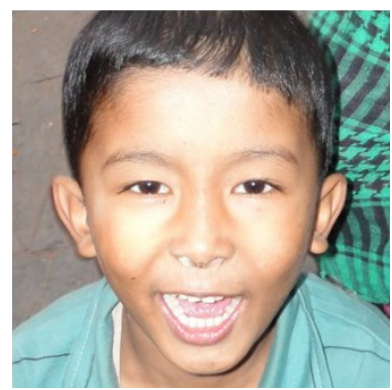 |

Figure D2. Visual examples for large SR: 32 to 512

| LR image<br>(64x64) | Bilinear LR image<br>(512x512) | SR image from DTLS<br>(512x512) | Ground Truth<br>(512x512) |
|---|---|---|---|
| 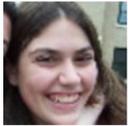 | 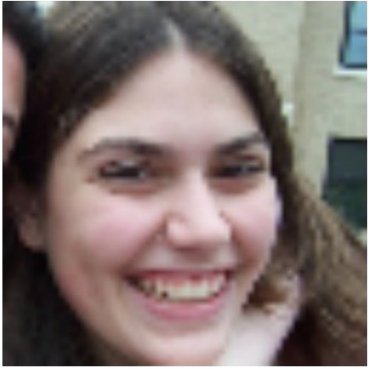 | 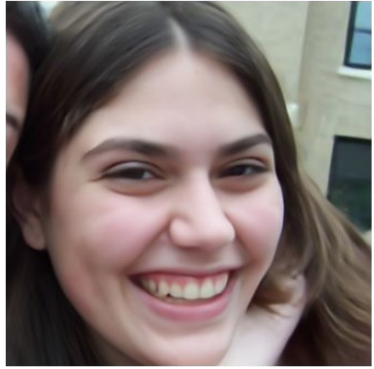 | 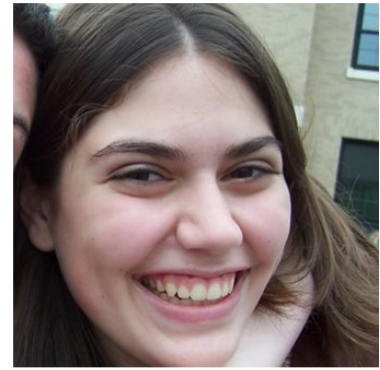 |
| 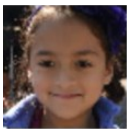 | 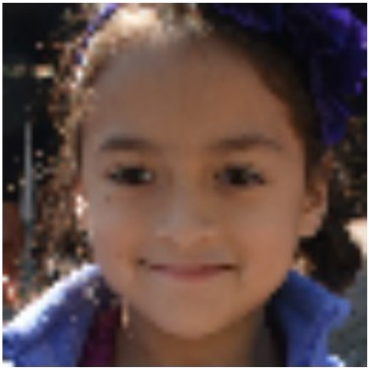 | 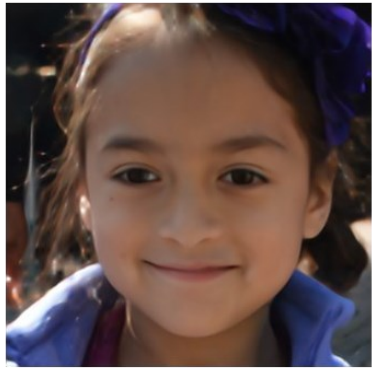 | 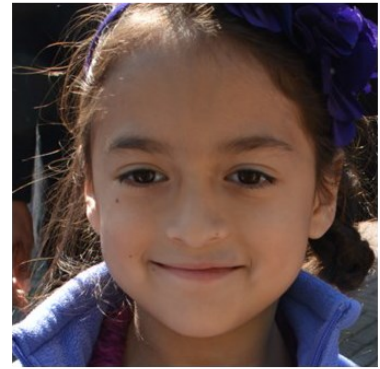 |
| 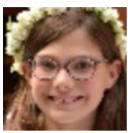 | 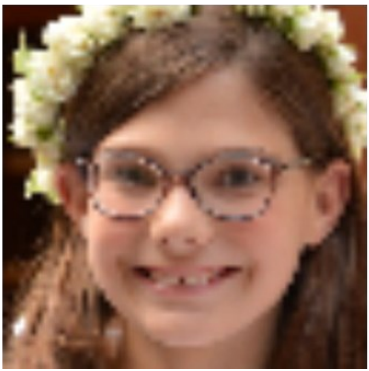 | 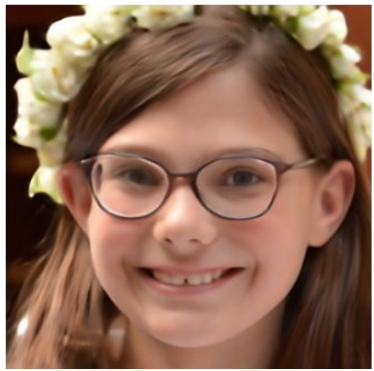 | 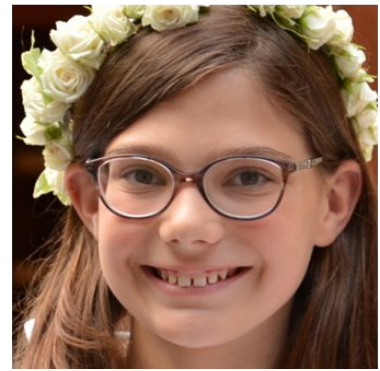 |

Figure D3. Visual examples for 64 to 512 SR

| LR image (64x64) | Bilinear LR image (512x512) | SR image from DTLS (512x512) | Ground Truth (512x512) |
| --- | --- | --- | --- |
| 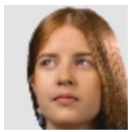 | 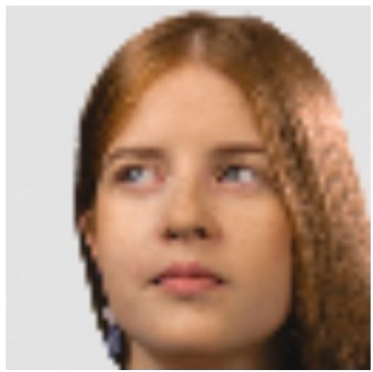 | 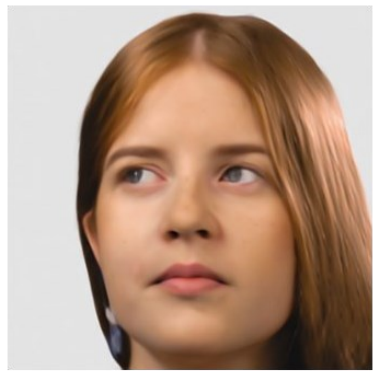 | 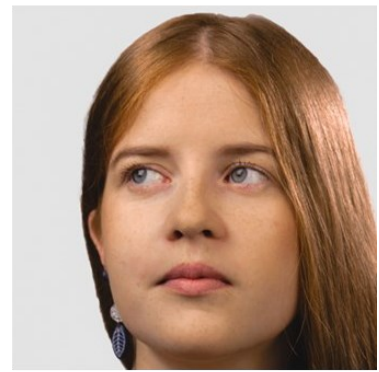 |
| 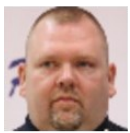 | 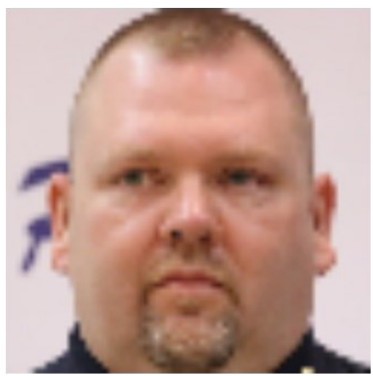 | 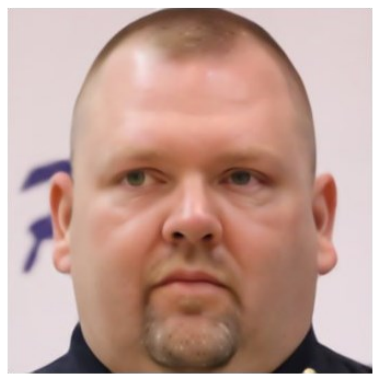 | 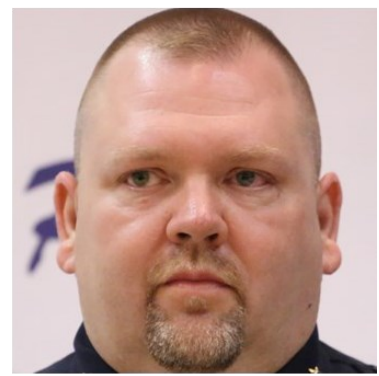 |
| 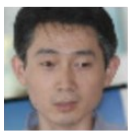 | 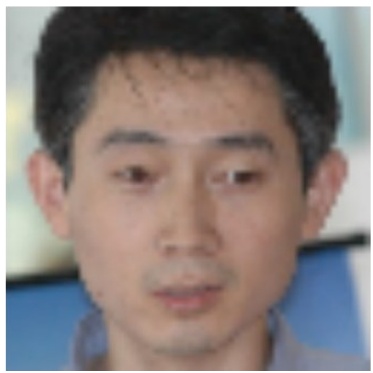 | 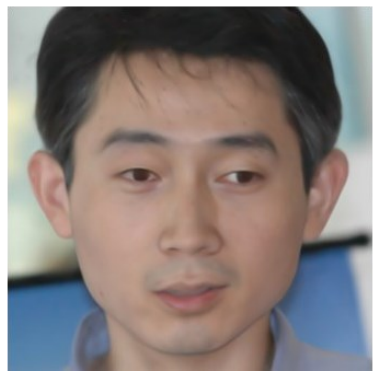 | 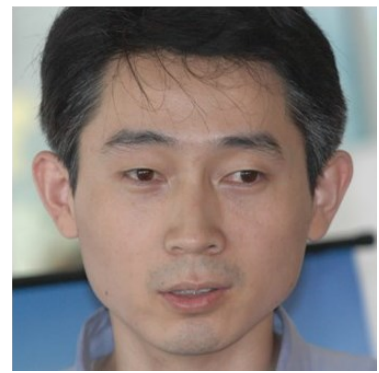 |

Figure D4. Visual examples for 64 to 512 SR

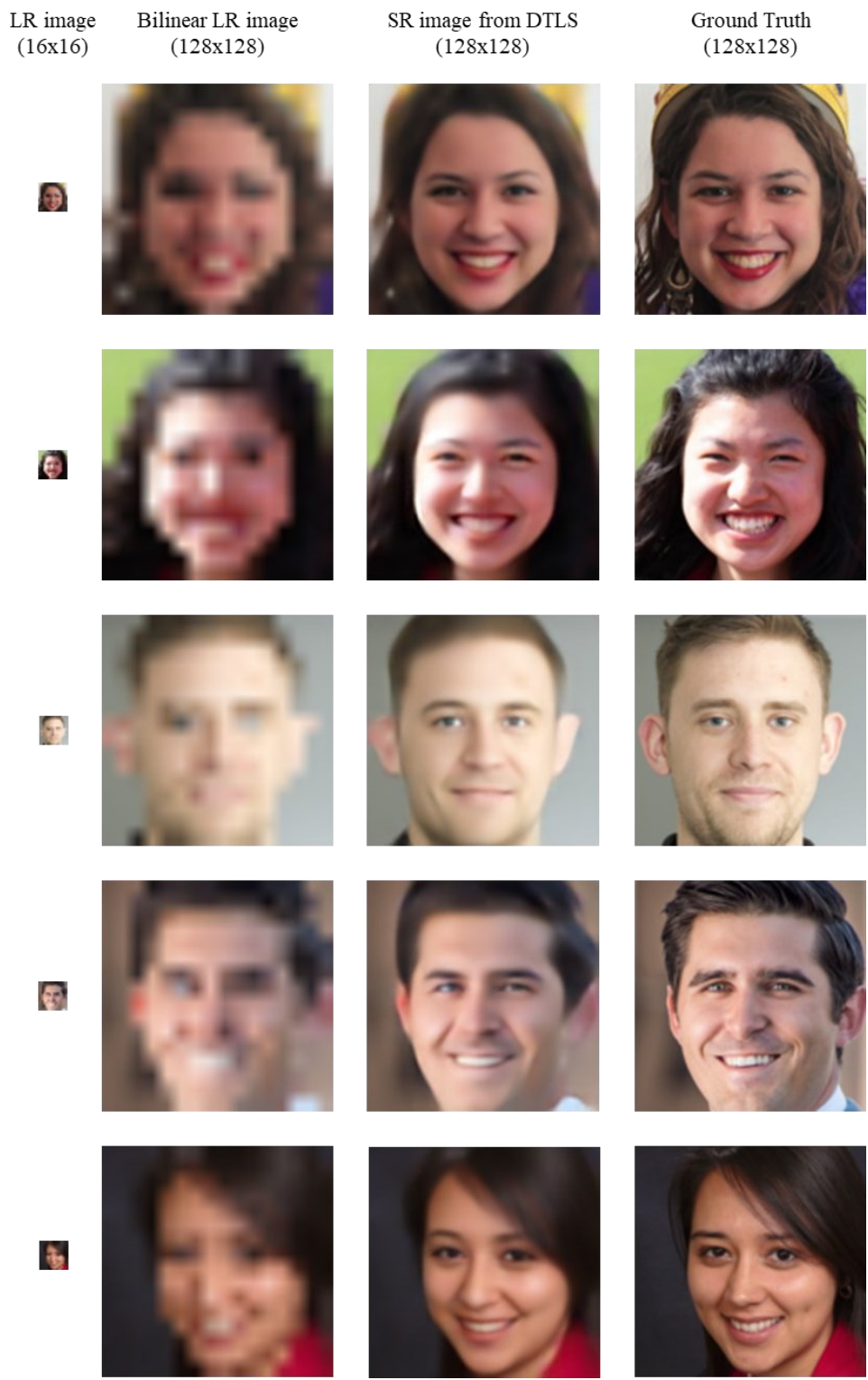

Figure D5. Visual examples of faint input for 16 to 128 SR

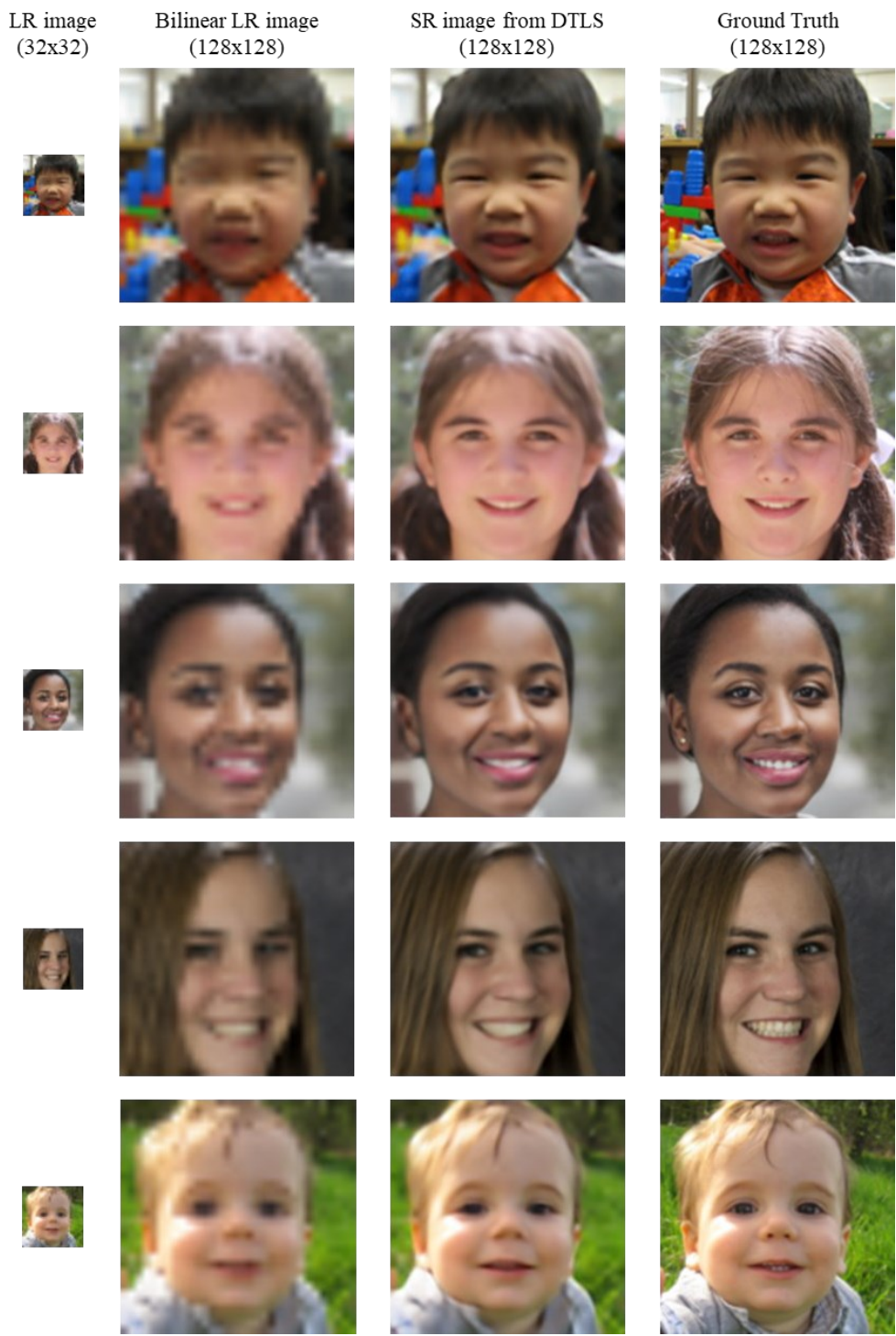

Figure D6. Visual examples for 32 to 128 SR

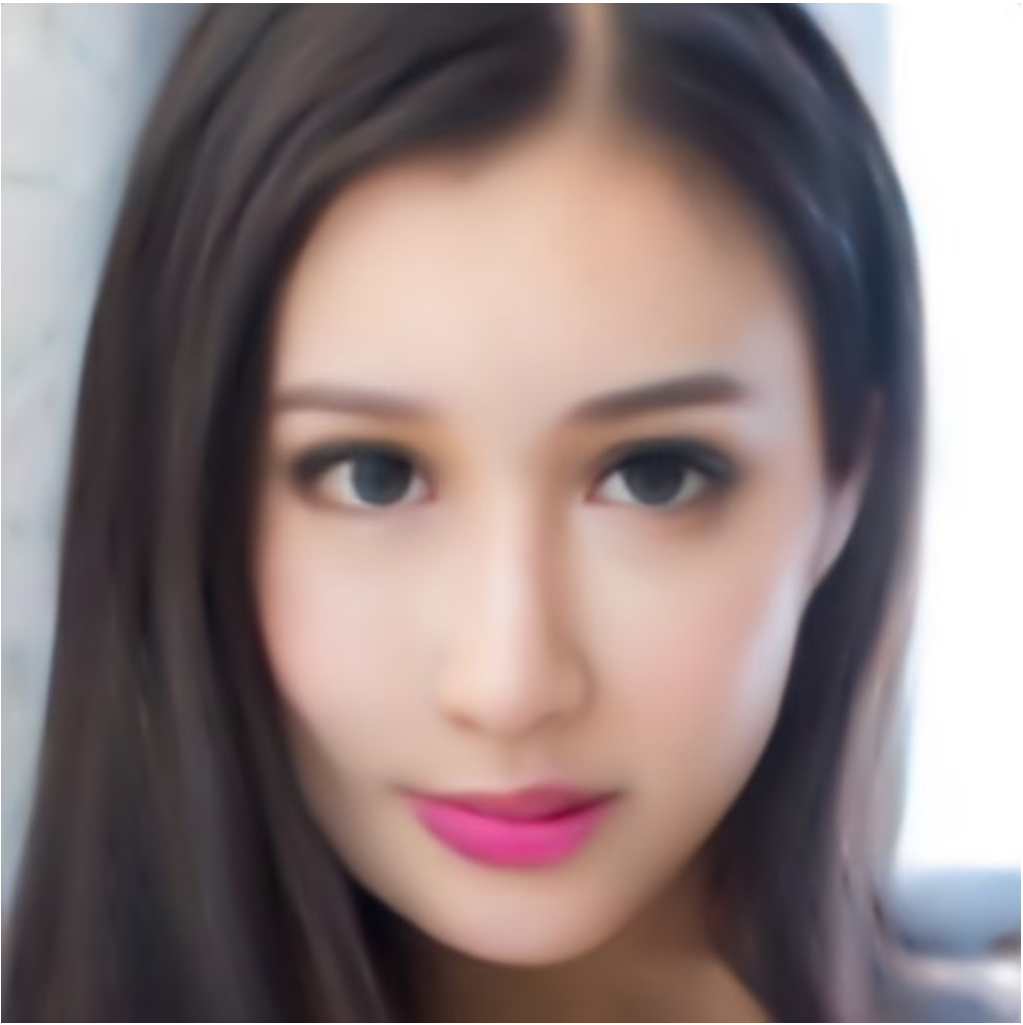

Figure D7. Big Size visual examples for 32 to 512 SR

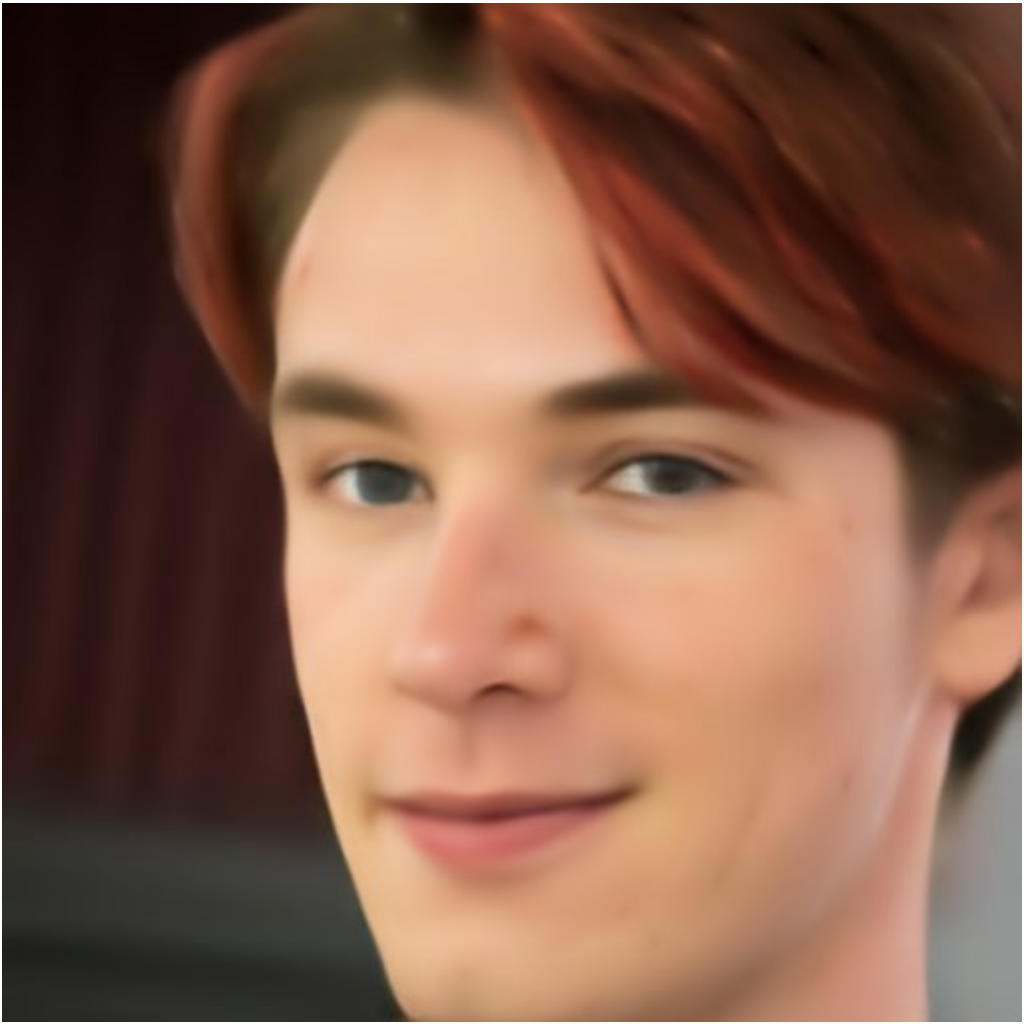

Figure D8. Big Size visual examples for 32 to 512 SR

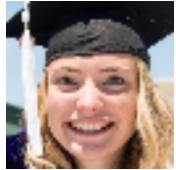

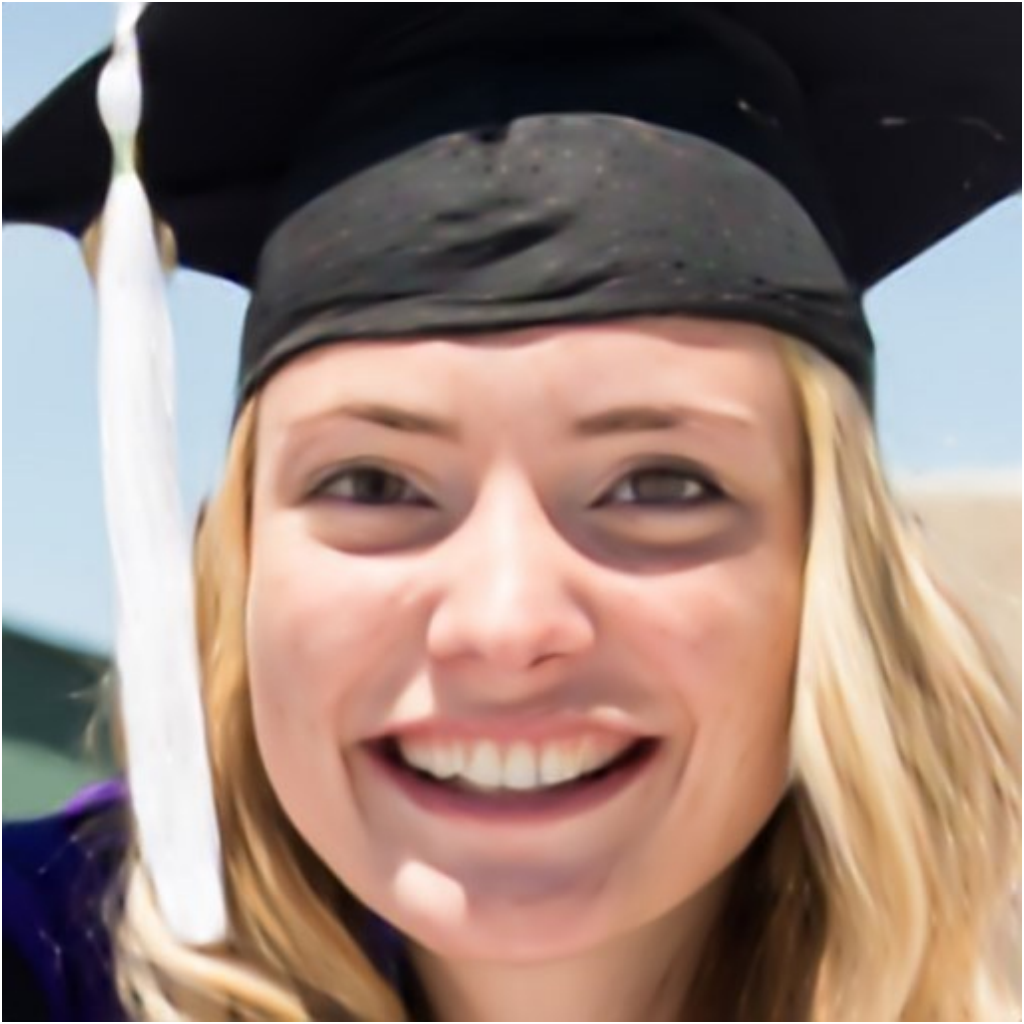

Figure D9. Big Size visual examples for 64 to 512 SR

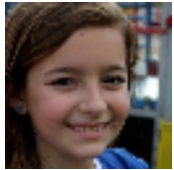

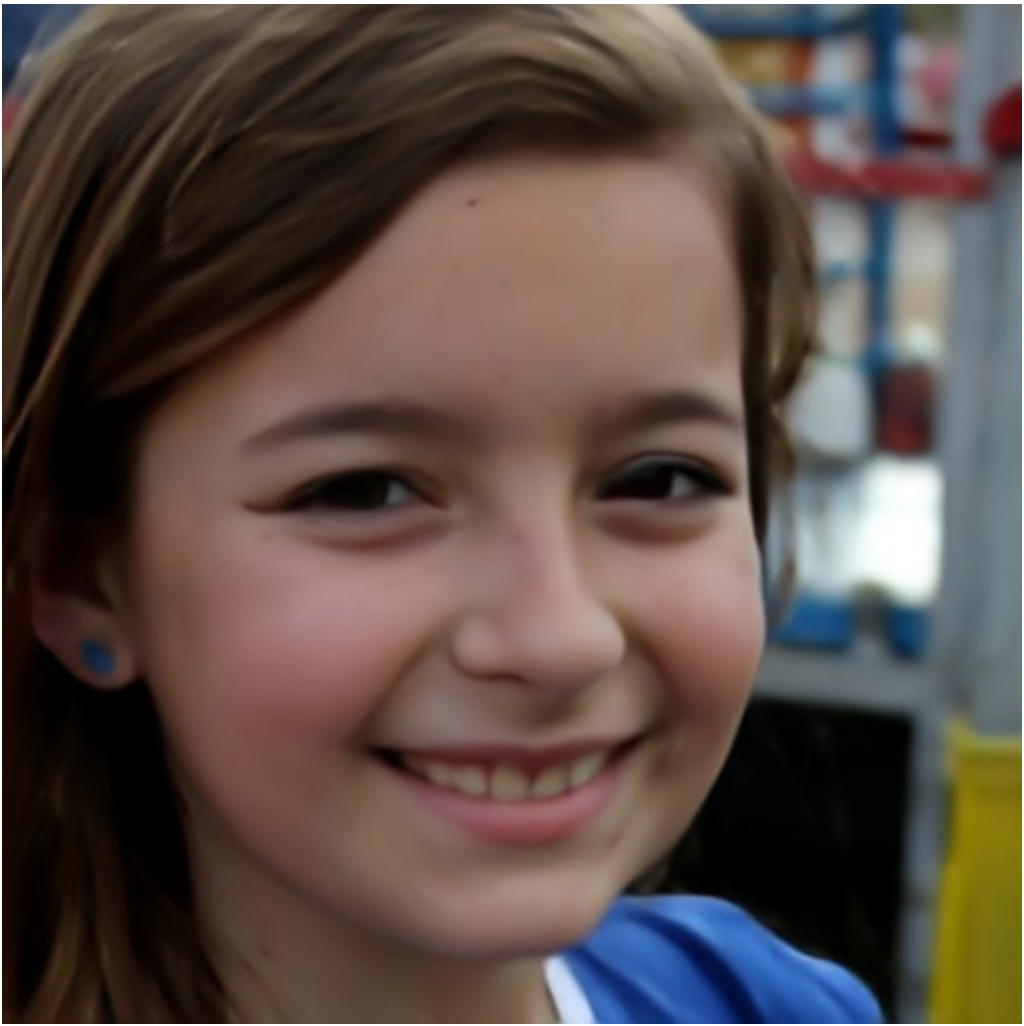

Figure D10. Big Size visual examples for 64 to 512 SR

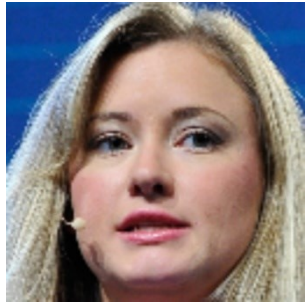

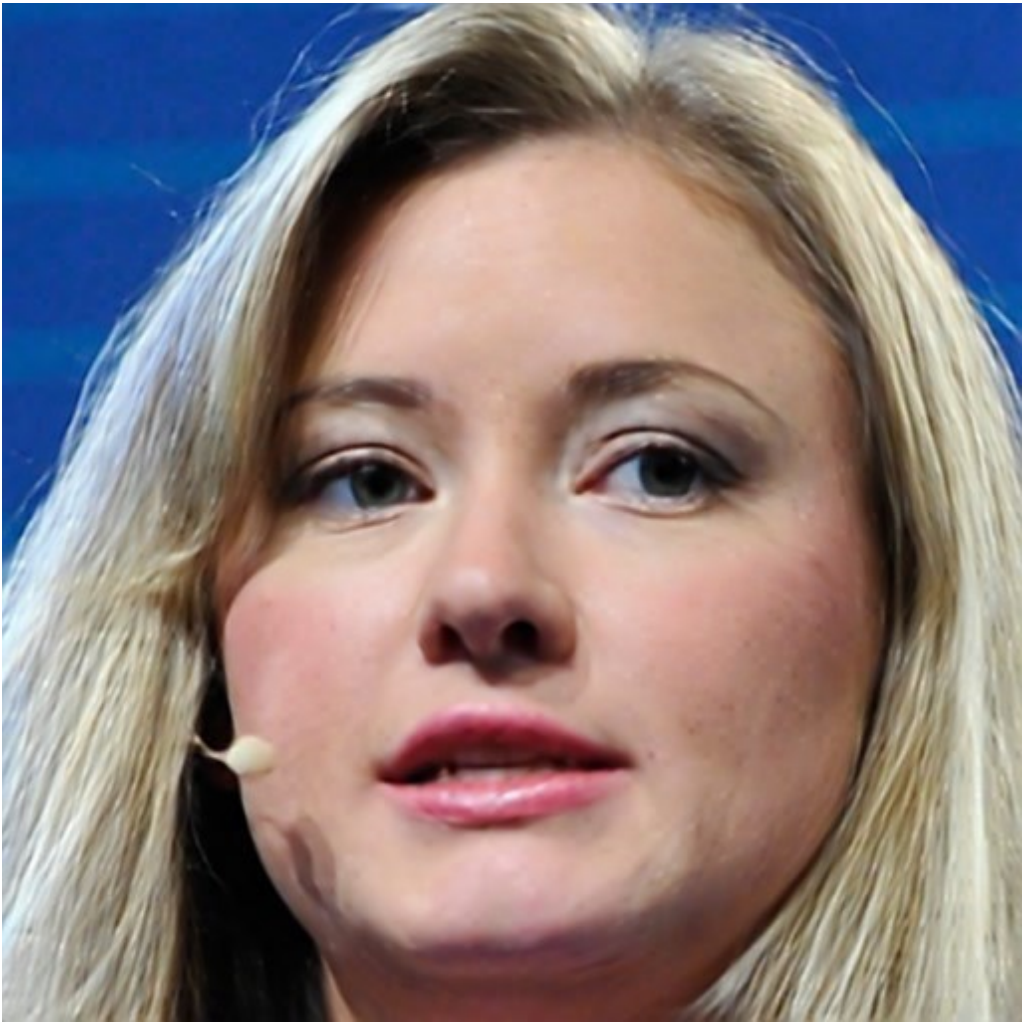

Figure D11. Big Size visual examples for 128 to 512 SR

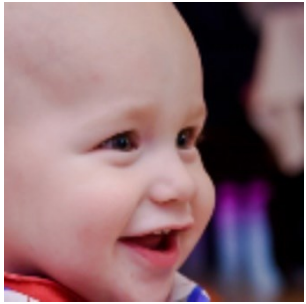

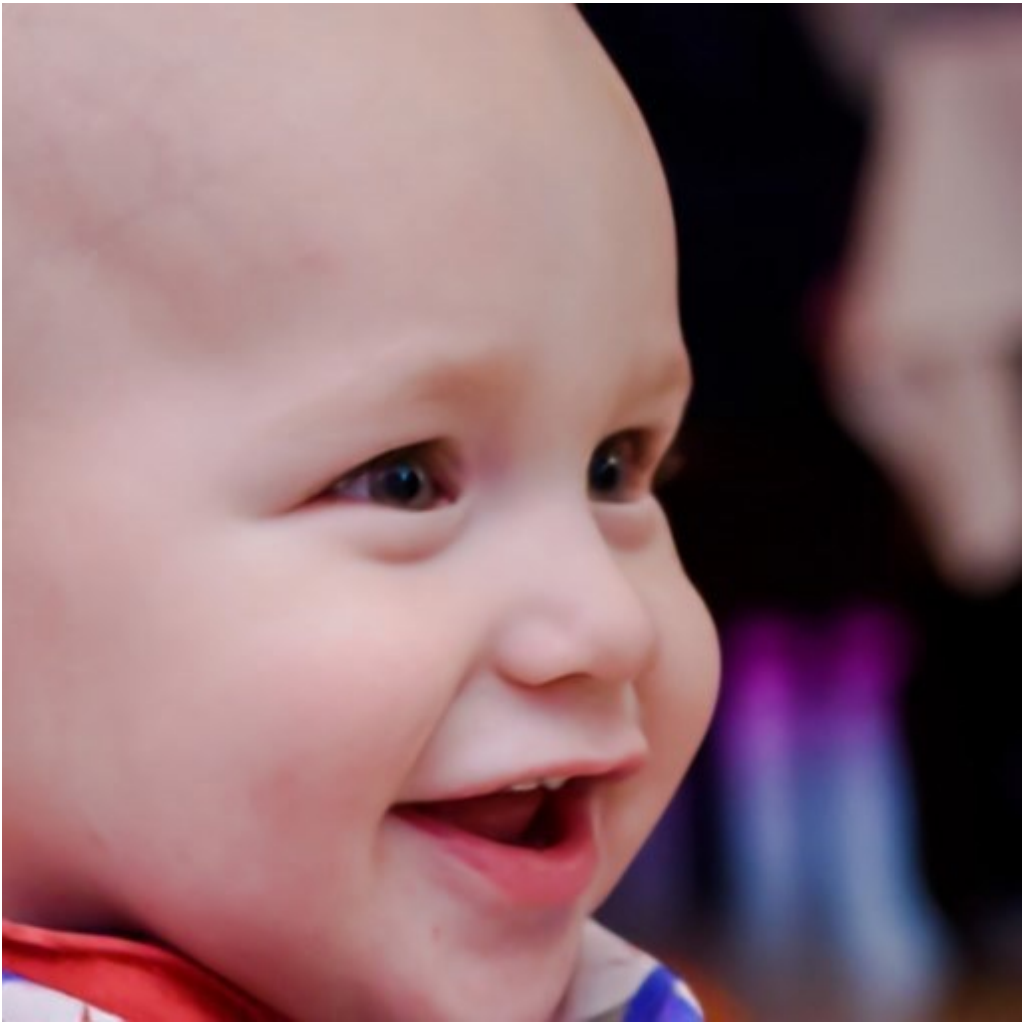

Figure D12. Big Size visual examples for 128 to 512 SR

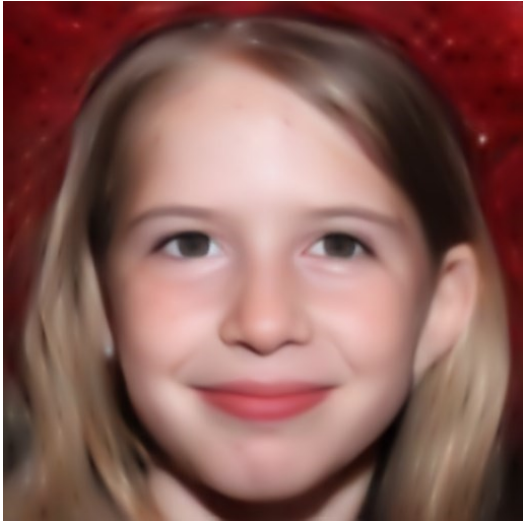
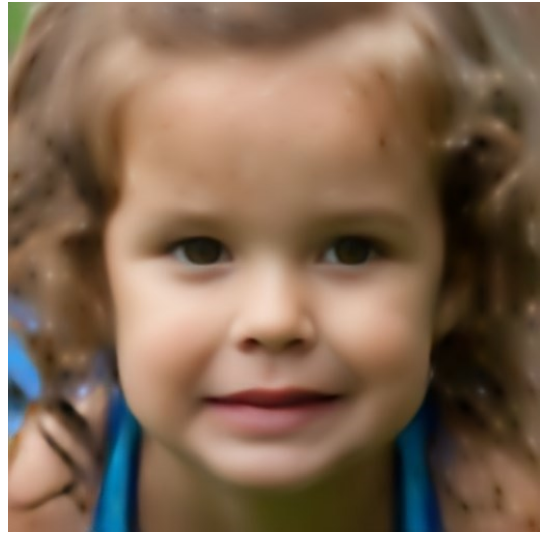

Fig.D13a　　　　　　　　　　　　　　　　　Fig.D13b

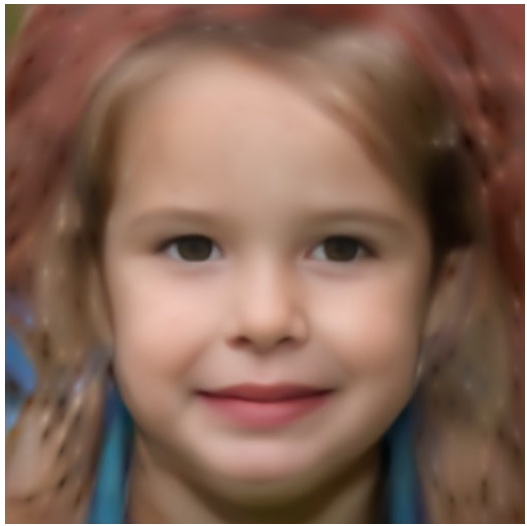

Fig.D13c: Image Generated by DTLS

Figure D13. Demonstration of Generative Capability: Fig.D13c is a generated image using our proposed approach (DTLS) from a mix of the latent vectors of D13a and D13b. This illustration shows that DTLS possesses good generative power.